\documentclass[%
reprint,
prd,
amsmath,amssymb,aps,
]{revtex4-2}
\usepackage [latin1]{inputenc}
\usepackage{graphicx}
\usepackage{dcolumn}
\usepackage{bm}
\usepackage{amsmath}

\usepackage{ifthen} 
\usepackage{fp,calc}
\usepackage{lastpage}
\IfPackageLoadedTF{color}{}{\usepackage{color}}

\usepackage{fancyhdr}
\usepackage{adjustbox}
\usepackage[english]{babel}

\IfPackageLoadedTF{xcolor}{}{\usepackage[svgnames, table]{xcolor}}
\usepackage{layout}
\usepackage{enumitem}
\usepackage[normalem]{ulem}
\usepackage{parskip}
\usepackage{multirow}
\usepackage{comment}
\usepackage{booktabs}
\usepackage[flushleft]{threeparttablex}

\newboolean{articletitles}
\setboolean{articletitles}{true} 
\newboolean{allauthors}
\setboolean{allauthors}{false} 

\providecommand\mnras{Mon.\ Not.\ Roy.\ Astron.\ Soc. }
\providecommand\aap{Astron.\ Astrophys. }
\providecommand\prd{Phys.\ Rev.\ D}

\providecommand\jcap{JCAP }

\providecommand\apj{Astrophys.\ J. }
\providecommand\aj{Astron.\ J. }

\usepackage{hyperref}

\newcommand{\hJ}{\hat{J}}
\newcommand{\hf}{\hat{f}}
\newcommand{\HH}{\mathcal{H}}
\newcommand{\bV}{\mathbf{V}}
\newcommand{\bk}{\mathbf{k}}
\newcommand{\NN}{\mathcal{N}^{\rm grow}}

\begin{document}

\title{A novel test of gravity: Does spacetime geometry track matter density?}

\author{Camille Bonvin}
\email{camille.bonvin@unige.ch}
\author{Nastassia Grimm}
\email{nastassia.grimm@unige.ch}
\author{Isaac Tutusaus}
\email{isaac.tutusaus@irap.omp.eu}
\affiliation{%
D\'epartement de Physique Th\'eorique and Center for Astroparticle Physics, Universit\'e de Gen\`eve, Quai E. Ansermet 24, CH-1211 Geneva 4, Switzerland}%
\affiliation{
 Institut de Recherche en Astrophysique et Plan\'etologie (IRAP), Universit\'e de Toulouse, CNRS, UPS, CNES, 14 Av.~Edouard Belin, 31400 Toulouse, France}%

\date{\today}

\begin{abstract}
We propose a novel test of gravity that combines galaxy clustering with gravitational lensing. In general relativity, the evolution of matter density fluctuations and of the Weyl potential -- the sum of spatial and temporal distortions of the geometry -- are governed by the same growth function. In contrast, alternative theories of gravity that modify the relation between geometry and matter content can lead to differences in these two growths. Exploiting a recent method to directly measure the Weyl potential, we construct a null test that deviates from zero if and only if there is a mismatch between the growth rate of density and that of geometry distortions. 
We show that changes in the background expansion due to alternative dark energy models and additional forces in the dark matter sector induce no deviations in this test, making it a robust probe for detecting departures from general relativity. Applying the test to current data, we find no evidence of deviation. From an initial $z_*=10$  to $z\sim 0.5$, we constrain the evolution of the Weyl potential to track that of the density to within 33\%.  Combining stage-IV surveys will improve the precision across a broad redshift range, limiting differences between the two evolutions to below $2-4\%$. 
\end{abstract}

\maketitle

Two fundamental mysteries currently challenge our understanding of the Universe. First, the mechanism driving its accelerated expansion is still unknown~\cite{SupernovaSearchTeam:1998fmf,Perlmutter_1999}. Second, the nature and fundamental properties of dark matter continue to elude us~\cite{Bertone:2016nfn,Blumenthal:1984bp,Davis:1985rj}. To model these two mysteries, cosmologists have built the $\Lambda$CDM model. In this model, the accelerated expansion of the Universe is caused by a cosmological constant, and dark matter is a cold non-interacting particle. While the $\Lambda$CDM model can explain the vast majority of our observations, various tensions have emerged over the past few years, such as the $H_0$ tension~\cite{Verde:2019ivm,DiValentino:2021izs}, the $\sigma_8$~\cite{2021APh...13102604D,Nunes:2021ipq} tension, the kinematic dipole tension~\cite{Secrest:2022uvx,Dalang:2021ruy}, and the evidence for a dynamical dark energy component~\cite{DESI:2024mwx,DESI:2025zgx,Berti:2025phi}. These tensions could be due to unaccounted systematic  effects in the data, but they could also indicate that the $\Lambda$CDM model is not the correct theory to describe our Universe. In this context, it is of crucial importance to test the two pillars of $\Lambda$CDM, namely the theory of General Relativity (GR), which governs the evolution of the Universe, and the matter and energy content of the model, i.e.\ the existence of cold dark matter and of the cosmological constant.

The Universe and its large-scale structure provide a remarkable laboratory to test these two paradigms. Such studies are however plagued by two difficulties. First, there exists a plethora of models beyond $\Lambda$CDM, in the modified gravity sector, in the dark energy sector and in the dark matter sector, and it has become unfeasible to confront them with observations one by one. Second, there are strong observational degeneracies between different models. For example, having a fifth (non-gravitational) force acting on dark matter, or modifying gravity by changing Poisson's equation have exactly the same impact on the growth of structure observed through galaxy surveys~\cite{Castello:2022uuu,Bonvin:2022tii,Castello:2024jmq}. In this context, it is crucial to develop observational tests that isolate and target specific physical features of models beyond $\Lambda$CDM. 

A variety of such tests have been proposed in the past decades, for example tests of the distance duality relation~\cite{Bassett:2003vu,Avgoustidis:2010ju,Keil:2025ysb,Yang:2025qdg}, of the cosmological principle~\cite{Secrest:2020has,Nadolny:2021hti,Secrest:2022uvx}, of the weak equivalence principle~\cite{Bonvin:2018ckp,Castello:2024lhl}, of the scale-invariance of the growth of structure~\cite{Franco:2019wbj}, or consistency tests of the $\Lambda$CDM model, e.g.\ through evolution equations~\cite{Arjona:2024maz} or through the $E_G$ statistics~\cite{Zhang:2007nk,Alam:2016qcl,Amon2018MNRAS.479.3422A, Reyes:2010tr, Blake:2016fcm, Pullen2016, Wenzl:2024sug, Abidi_2023_10419,Grimm:2024fui}. In this work, we introduce a novel test designed to assess whether the evolution of the Universe's geometry follows that of its matter density. In GR, the two are linked through Einstein's equations, but this relationship is typically altered in modified gravity theories. We therefore build a null test that deviates from zero if and only if there is a mismatch between the growth rate of the density and that of the geometry. This null test combines measurements of the Weyl potential, using a recent method that enables its reconstruction in redshift bins~\cite{Tutusaus:2023aux}, with measurements of the growth rate of structure from redshift-space distortions~\cite{Kaiser:1987qv,Hamilton:1995px}.

We show that only modifications of gravity lead to a deviation of the null test from zero. Even more notably, such deviations occur only within a specific class of modified gravity theories: those that alter the propagation of light in the Universe.
Theories of gravity that only modify the motion of massive objects, like galaxies, do not generate a non-zero null test. The null test is also insensitive to variations in the background evolution, and is therefore not affected by dark energy models beyond the cosmological constant. Finally, the null test is insensitive to the presence of a fifth force acting on dark matter. While such a force modifies the growth of structure~\cite{Castello:2022uuu}, it preserves the link between density and geometry, keeping the null test at zero. As a consequence, the null test specifically targets deviations in the gravitational sector, while remaining insensitive to modifications in the cold dark matter component and in the background evolution. This makes this test highly complementary to other tests that are affected by any possible deviation from the $\Lambda$CDM model, independently on their origin.

Applying the null test to current data, we obtain a precision of $0.11-0.26$ depending on redshift, and find no evidence for deviations from zero. We further show that combining stage-IV surveys will boost the precision of the test by a factor of $20-50$ across a wide redshift range.

\emph{Formalism}: We describe our Universe as a homogeneous and isotropic background with scalar perturbations. The geometry is encoded in the perturbed Friedmann-Lema\^itre-Robertson-Walker (FLRW) metric
\begin{align}
\mathrm ds^2=a^2(\tau)\left[-(1+2\Psi(\mathbf{x},\tau))\mathrm d\tau^2+ (1-2\Phi(\mathbf{x},\tau))\mathrm d\mathbf{x}^2\right]\,,\nonumber
\end{align}
where $a$ is the scale factor which depends on conformal time $\tau$, and $\Psi$ and $\Phi$ are the two gravitational potentials. Matter variables are similarly split into a background and perturbations. The total matter density is written as $\rho=\bar{\rho}(1+\delta)$, where $\bar{\rho}$ is the background density and $\delta$ the density contrast. The velocity field vanishes in the homogeneous background, where galaxies follow the Hubble flow, but in the perturbed Universe, they acquire non-zero peculiar velocities, denoted by $\bV$.

In GR, linear density fluctuations grow at the same rate on all scales~\cite{Dodelson:2003ft}. Consequently, the evolution of $\delta$, from an initial redshift $z_*$ to a lower redshift $z$, can be encoded in a growth function $D_1$:
\begin{align}
\delta(\bk,z)=\frac{D_1(z)}{D_1(z_*)}\delta(\bk,z_*)\, .   
\end{align}
This relation holds for any $z$ and $z_*$ small enough for radiation to be negligible, typically $z,z_* \lesssim 10$. 
Einstein's equations directly link the growth of the Weyl potential $\Psi_W\equiv(\Phi+\Psi)/2$, to that of the density. Using Poisson's equation and the fact that in GR $\Psi=\Phi$, we obtain
\begin{align}
\label{eq:PsiW}
\Psi_W(\bk,z)=\frac{\HH^2(z)}{\HH^2(z_*)}\frac{D_1(z)}{D_1(z_*)}\Omega_{\rm m}(z)\Psi_W(\bk,z_*)\, ,   
\end{align}
where $\Omega_{\rm m}$ is the matter density parameter at redshift $z$. Hence $\delta$ and $\Psi_W$ both grow with the same function $D_1$.

If gravity is modified, the relation between the growth of the matter density and that of the Weyl potential is typically altered. In such theories, Poisson's equation is generally modified, and the two gravitational potentials $\Phi$ and $\Psi$ no longer coincide. As a result Eq.~\eqref{eq:PsiW} must be replaced by~\cite{Tutusaus:2022cab}
\begin{align}
\Psi_W(\bk,z)=\frac{\HH^2(z)}{\HH^2(z_*)}\frac{J(z)}{D_1(z_*)}\Psi_W(\bk,z_*)\, ,  \end{align}
where $J(z)$ is a function determined by the underlying theory of gravity and reduces to $D_1(z)\Omega_{\rm m}(z)$ in GR. 
The goal of this work is to construct a null test that deviates from zero whenever the growth of the density and that of the Weyl potential are not governed by the same function~$D_1$. 

\emph{The null test}: The matter density and the Weyl potential are not directly observable in large-scale structure surveys. However, such surveys can be used to measure two related functions. First, spectroscopic galaxy surveys provide measurements of the galaxy two-point correlation function in redshift space, from which one can extract the quantity $\hat{f}(z)\equiv f(z)\sigma_8(z)$, see e.g.~\cite{Percival:2008sh,2012MNRAS.423.3430B}. Here, $f(z)$ is the growth rate that governs the evolution of galaxy peculiar velocities, and $\sigma_8$ is the amplitude of matter clustering in spheres of radius 8\,Mpc/$h$. Using the continuity equation, the growth rate $f(z)$ is linked to  $D_1$ via $f(z)=\mathrm d\ln D_1/\mathrm d\ln a$.
Measuring $\hf$ thus directly probes the redshift evolution of $D_1$.
The second quantity that can be observationally accessed is $\hat{J}(z)$, which is related to the growth of the Weyl potential $J(z)$ via
\begin{align}
\label{eq:hatJ}
\hat{J}(z)=J(z)\frac{\sigma_8(z)}{D_1(z)}\, . 
\end{align}
As shown in~\cite{Tutusaus:2022cab,Tutusaus:2023aux}, the function $\hJ$ can be measured at the redshifts of the lenses by combining galaxy-galaxy lensing with galaxy clustering.

Since in GR, $\hJ$ reduces to $\hat{J}(z)=\Omega_{\rm m}(z)\sigma_8(z)$, its redshift derivative is directly linked to $\hf$. We therefore construct the following null test: 
\begin{align}
\label{eq:def_test}
\NN(z) \equiv \frac{\mathrm d}{\mathrm dz}\left(\frac{\hat{J}(z)}{\Omega_{\rm m}(z)} \right)+\frac{\hat{f}(z)}{1+z}\, ,  
\end{align}
which identically vanishes in GR~\footnote{Here we have used that at late time, once radiation is negligible, $D_1(z)/\sigma_8(z)$ is constant. In practice, both the residual radiation present at $z_*=10$ and the presence of massive neutrinos induce a small time evolution of this ratio. This results in a deviation of about 0.001 in $\NN$, which, as we will see, is 5 times smaller than the most stringent $1\sigma$ uncertainty from the coming generation of surveys.}. A deviation from zero in $\NN$ indicates that $\hJ$ does not evolve as $D_1$. This null test therefore provides a direct method for identifying a mismatch between the growth of the Weyl potential and that of the matter density perturbations.

$\NN$ can be measured across redshift by combining measurements of $\hf$ with the derivative of $\hJ$, which can be numerically inferred from discrete measurements of $\hJ$. $\NN$ also depends on the matter density parameter $\Omega_{\rm m}(z)$. In this work, we assume a $\Lambda$CDM evolution for the background, and infer $\Omega_{\rm m}(z)$ from its present-day value $\Omega_{{\rm m}0}$, which is measured jointly with $\hJ$ from gravitational lensing~\cite{Tutusaus:2023aux}. Alternatively, if one wishes to relax the $\Lambda$CDM assumption for the background, $\Omega_{\rm m}(z)$ can be independently reconstructed from Baryon Acoustic Oscillation and supernova measurements. 

Importantly, both $\hat{f}$ and $\hat{J}$ are measured in a model-independent way, without assuming any specific theory of gravity or dark matter model, and can thus consistently be used to probe models beyond $\Lambda$CDM. These measurements rely only on three assumptions: (1) that at high redshift $z_*$, the standard matter power spectrum predicted by GR and measured by Planck is recovered (this is motivated by the fact that no deviations have been observed at high redshift; note that this assumption automatically excludes early dark energy models from our test); (2) that the background expansion history effectively follows that of $\Lambda$CDM; and (3) that $\hat{f}$ and $\hat{J}$ are scale independent. None of these assumptions are fundamental, and the measurements can be performed without them~\cite{Amendola:2022vte,Schirra:2024rjq}. However, this introduces additional complexity, hence here we adopt these assumptions.

\begin{figure}
    \centering
    \includegraphics[width=.495\textwidth]{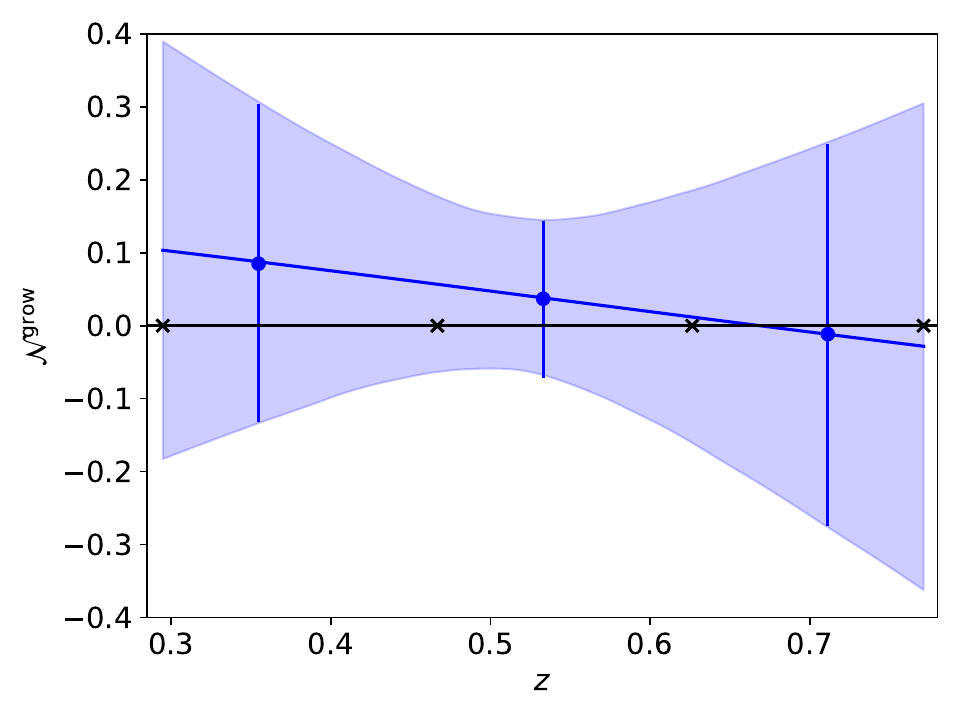}
    \caption{Measurements of $\NN$,  from current data sets,  together with the 1$\sigma$ uncertainties. We show the results at the three nodes (blue points) and the reconstructed $\NN$ over the whole redshift range. The black crosses indicate the position of the DES redshifts where $\hJ$ is measured.} \label{fig:current}
\end{figure}

\emph{Measurements}: To measure $\NN$, we use 22 measurements of $\hf$ between $z = 0.001$ and $z = 1.944$, from different galaxy surveys~\cite{Howlett:2017asq,Huterer:2016uyq,Hudson:2012gt, Turnbull:2011ty,Davis:2010sw,Song:2008qt,Blake:2013nif,eBOSS:2020yzd,Blake:2012pj,Pezzotta:2016gbo,Okumura:2015lvp,eBOSS:2018yfg}\footnote{Note that we do not include the recent measurements of $\hat{f}$ from DESI~\cite{DESI:2024jis}, since they may be correlated with the other measurements, and we do not have access to the covariance. As we will see the uncertainties on $\NN$ are dominated by uncertainties on the Weyl potential, hence adding DESI measurements is 
not expected to modify the results.}. The measurements with their uncertainties are listed in Table~I of~\cite{Grimm:2024fui}. We combine them with the measurements of $\hJ$ from DES at four redshifts, $z_{\rm DES}\in \{
0.295, 0.467, 0.626, 0.771\}$, obtained in~\cite{Tutusaus:2023aux}. We consider the baseline case presented there, with Planck priors for the cosmological parameters (first column in Table~I of~\cite{Tutusaus:2023aux}).
We construct $\NN$ on a set of redshift nodes, in the range where both $\hJ$ and $\hat{f}$ are measured. To obtain $\frac{\mathrm d}{\mathrm dz}\left(\frac{\hat{J}(z)}{\Omega_{\rm m}(z)} \right)$ at the nodes, we use the MCMC chain from~\cite{Tutusaus:2023aux}, where both $\hJ$ and the cosmological parameters are varied. At each step of the chain we compute $\hat{J}(z_{\rm DES})/\Omega_{\rm m}(z_{\rm DES})$. We treat the values of this ratio at the nodes as free parameters, and we determine them by minimizing the difference between the interpolating curve (using cubic spline) that passes through the nodes and the values from the chain at $z_{\rm DES}$. This properly accounts for the correlations between $\hJ$ and $\Omega_{{\rm m } 0}$. To choose the number of redshift nodes, we minimize the Akaike information criterion~\cite{Liddle:2007fy} (AIC) in the reconstruction of $\hJ$. 
We find that the AIC is similar for three and four nodes, and we choose three nodes as the baseline case with $z_{\rm nodes}\in \{0.35, 0.53, 0.71 \}$. In Fig.~\ref{fig:Jhat} of Supplemental Materials, we show the reconstructed $\hJ$ and its derivative. We also explore an alternative method, computing the derivative of $\hJ/\Omega_{\rm m}$ directly, by interpolating between the DES redshifts at each step of the chain.

For the reconstruction of $\hf$, we select four redshift nodes -- again chosen to minimize the AIC -- equally spaced between $z = 0.001$ and $z = 1.944$. We determine the values of $\hf$ and the covariance at these nodes through interpolation and minimization.

We then construct $\NN$ along with its uncertainty, by generating 10'000 samples of $\frac{\mathrm d}{\mathrm dz}\left(\frac{\hat{J}(z)}{\Omega_{\rm m}(z)} \right)$ and of $\hf/(1+z)$. The mean and 1$\sigma$ uncertainty of $\NN$ are plotted in Fig.~\ref{fig:current} and listed in Table~\ref{tab:Ncurrent} of Supplemental Materials. We see that $\NN$ is consistent with zero, showing no evidence of departure from GR. We also show the reconstructed mean and 1$\sigma$ uncertainties over the DES redshift range. Note that in practice, the reconstruction with three nodes is equivalent to fitting a second-order polynomial, hence the position of the nodes has no impact on $\NN$. We see that the uncertainty increases near the edges of the redshift range, due to the fact that $\NN$ depends on the derivative of $\hJ$, which is less well constrained near the edges.  Comparing with the interpolating method of Supplemental Materials (see Fig.~\ref{fig:Ninterpolation}), we see that in that case $\NN$ fluctuates around zero. This is due to the interpolation of $\hJ$, as can be seen from Fig.~\ref{fig:interpolation}. The results are however also compatible with GR (at $1.5\sigma$), and the uncertainties are similar, albeit slightly larger than for three nodes.

To better interpret the meaning of the uncertainties, $\sigma_{\NN}$, we rewrite the null test in terms of the growth function of the density, $D_1$, and the growth function of the Weyl potential, that we call $\tilde{D}_1$. We find that a deviation of $\NN$ from zero by an amount $\sigma_{\NN}$ leads to the relation
\begin{align}
&\frac{\mathrm d}{\mathrm dz}\left(\frac{\tilde{D}_1(z)}{\tilde{D}_1(z_*)}\right)=\frac{\mathrm d}{\mathrm dz}\left(\frac{D_1(z)}{D_1(z_*)}\right)\left(1 + \rm{dev} \right)\, ,    \end{align}
\begin{align}
\mbox{with} \quad\quad |{\rm dev}|=
\frac{\sigma_{\NN}(1+z)}{\hf(z)}\, . 
\end{align}
We obtain ${\rm dev}=\{0.58, 0.33, 0.92\}$. This means, e.g., that between $z_*=10$ and $z=0.53$, the evolution of $\Psi_{W}$ cannot differ from that of $\delta$ by more than 33\%.

\begin{figure}
    \centering
    \includegraphics[width=.495\textwidth]{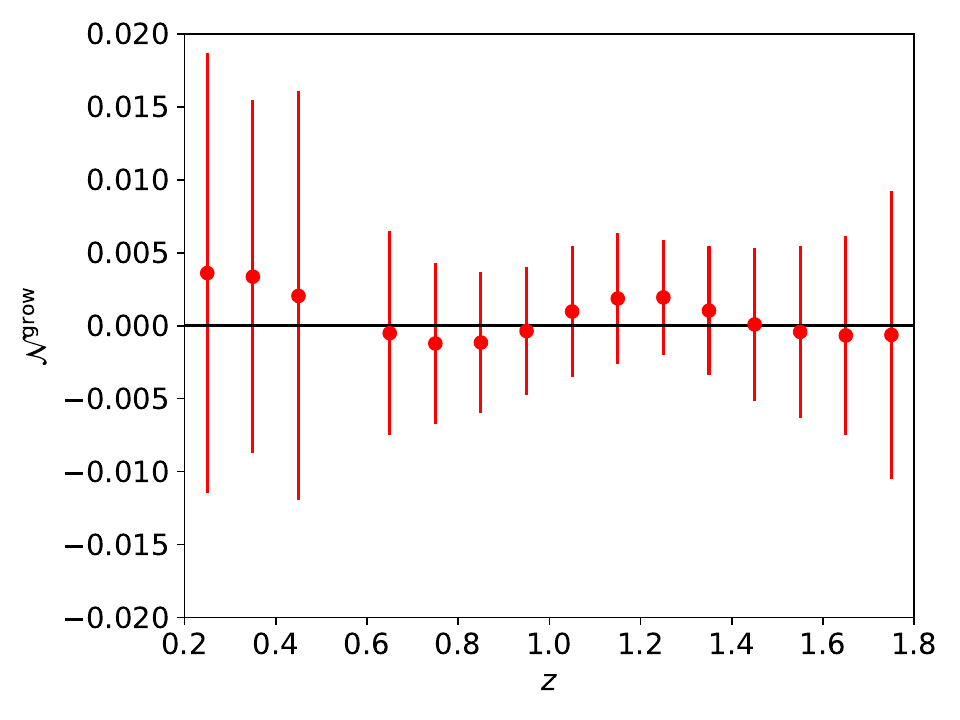}
    \caption{Forecasted mean and 1$\sigma$ uncertainties for $\NN$ obtained by combining LSST and DESI. }\label{fig:future}
\end{figure}

\emph{Forecasts for DESI\,\footnote{\url{https://www.desi.lbl.gov}} and LSST\,\footnote{\url{https://rubinobservatory.org}}}: To assess the improvement expected from stage-IV surveys, we forecast the precision on $\NN$ achievable by combining LSST- and DESI-like data. For $\hJ$, we adopt the LSST forecast from~\cite{Tutusaus:2022cab}, which provides $\hJ$ and its covariance matrix in ten redshift bins between $z=0.25$ and $z=2.1$. For $\hf$, we assume measurements at 16 redshifts between $z=0.25$ and $z=1.85$. The $1\sigma$ uncertainties correspond to forecasts for DESI's final data release, taken from Tables 2.3 and 2.5 of~\cite{DESI:2016fyo}, assuming that DESI will observe approximately 30 million galaxies over 14'000 square degrees.

To infer $\frac{\mathrm d}{\mathrm dz}\left(\frac{\hat{J}(z)}{\Omega_{\rm m}(z)} \right)$ we select six redshift nodes -- chosen to minimize the AIC -- within the LSST redshift range, and treat the values of $\hJ$ at these nodes as free parameters. As before, these parameters are determined through interpolation and minimization. To properly infer the covariance between $\hJ$ at the six nodes and $\Omega_{{\rm m}0}$, we numerically compute the Jacobian that transforms the original Fisher matrix -- containing $\hJ$ at the LSST redshifts along with the cosmological and nuisance parameters from~\cite{Tutusaus:2022cab} -- into a new Fisher matrix defined in terms of $\hJ$ at the six nodes and the same set of cosmological and nuisance parameters. We then draw 10'000 samples from the resulting joint covariance matrix of $\hJ$ and $\Omega_{{\rm m}0}$, and use spline interpolation to obtain $\frac{\mathrm d}{\mathrm dz}\left(\frac{\hat{J}(z)}{\Omega_{\rm m}(z)} \right)$ at the DESI redshifts. These results are combined with independent samples of $\hf$, to construct $\NN$ at those redshifts. The results are plotted in Fig.~\ref{fig:future} and listed in Table~\ref{tab:Nfuture} of Supplemental Materials. Over a wide redshift range, $\NN$ can be constrained with a precision of 0.005, i.e.\ $20-50$ times better than what is currently achievable. This translates into a sensitivity to differences between the evolution of $\Psi_W$ and that of $\delta$ at the level of $2-4\%$.

\emph{Deviations in the null test}: $\NN$ is measured without assuming a specific theory of gravity. If a deviation is observed, it can then be used to constrain theories beyond GR. As an illustrative example, we consider the standard phenomenological $\mu-\Sigma$ parametrization of modified gravity~\cite{Amendola:2007rr,Bertschinger:2008zb,2010PhRvD..81j4023P}. As shown in Supplemental Materials, such modifications to GR change the motion of matter and thus the growth function $D_1$ through the function $\mu(z)$ (see Eq.~\eqref{eq:deltaevol_mod}); and they also change
the relation between $\Psi_W$ and $\delta$ through the function $\Sigma(z)$:
\begin{align}
k^2\Psi_W(k,z)=-\frac{3}{2}\HH^2\Sigma(z)\delta(k,z)\, . \label{eq:Psi_Sigma} 
\end{align}
The change in $D_1$ due to $\mu\neq 1$ does not alter $\NN$, since the same modified $D_1$ governs both the growth of $\Psi_W$ and that of $\delta$, preserving their relationship. On the contrary, $\Sigma\neq 1$ directly impacts the propagation of light, leading to $\hJ=\Omega_{\rm m}(z)\Sigma(z)\sigma_8(z)$ and thus $\NN\neq 0$. $\NN$ therefore allows us to target one specific fundamental aspect of gravity: its effect on light propagation, without being affected by changes in the motion of matter. We also see that $\NN$ cannot be zero if $\Sigma\neq 1$: even if $\mu$ and $\Sigma$ are adjusted to keep $\hJ/\Omega_{\rm m}$ unchanged, i.e.\ $\hJ/\Omega_{\rm m}=\sigma^{\Lambda{\rm CDM}}_8$, the null test would show a deviation since $\hf\propto  \mathrm d\sigma_8/\mathrm dz\neq \mathrm d\sigma^{\Lambda{\rm CDM}}_8/\mathrm dz$. The only exception that would not be detected in $\NN$ is a change in the growth of $\hJ$ that would leave the growth rate (i.e.\ the slope) unchanged. Finally, we see that changes in the background evolution do not impact $\NN$. Modifying the background would affect $D_1$, but this modification would not alter the relation between $\Psi_W$ and $\delta$, keeping $\NN=0$. As such $\NN\neq 0$ unequivocally signals a deviation in light propagation.

\begin{figure}[!t]
    \centering
    \includegraphics[width=.495\textwidth]{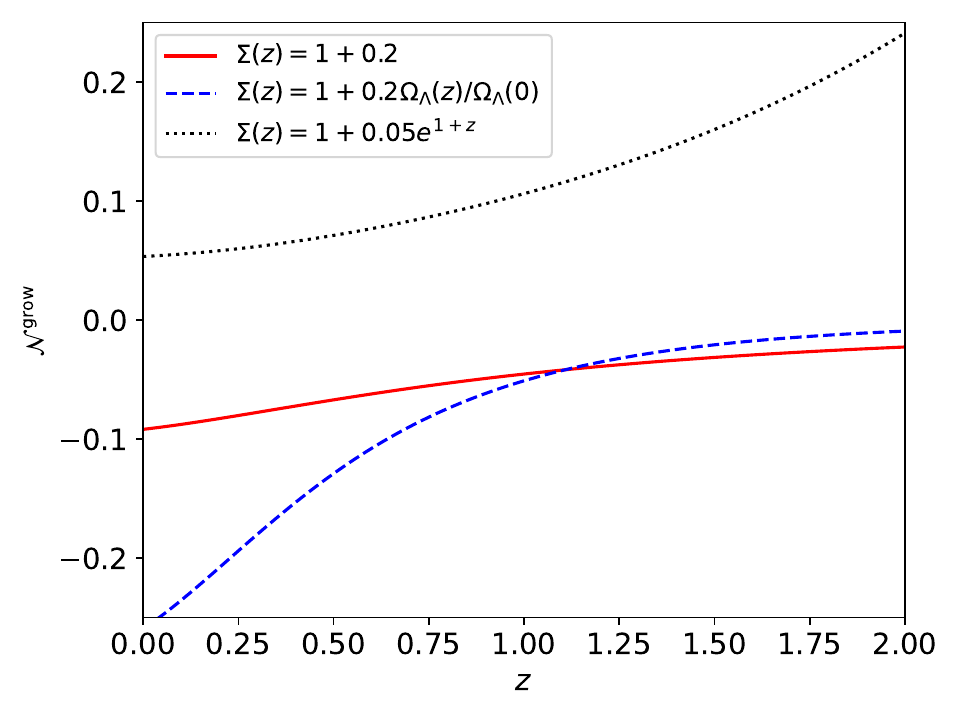}
    \caption{Deviations in $\NN$ induced by $\Sigma(z)\neq 1$, for three different redshift evolutions.}\label{fig:N_Sigma}
\end{figure}

Deviations in $\NN$ depend on the redshift evolution of $\Sigma(z)$. As examples, we consider the following choices, $\Sigma(z)=1+\Sigma_0 g(z)$,
with: (1) $g(z)=\Omega_\Lambda(z)/\Omega_\Lambda(z=0)$, which is the standard evolution used e.g.\ in~\cite{Planck:2015bue,DES:2022ccp,DESI:2024hhd}; (2) $g(z)={\rm cst}$ for $z\in [0,2]$ and zero elsewhere; and (3) $g(z)=\exp(1+z)$ for $z\in [0,2]$ and zero elsewhere. The first choice corresponds to a scenario in which deviations from GR intensify proportionally to the accelerated expansion. The second and third cases are not physically motivated, but simply chosen to assess the sensitivity of $\NN$ to very different types of deviations from GR.

\setlength{\tabcolsep}{0.8em} 
\begin{table}
\centering
\caption{Mean and 1$\sigma$ uncertainties for $\Sigma_0$ for different evolutions from current and future surveys.} \label{tab:Sigma0}
\begin{tabular}{@{}l r r@{}} 
\toprule
Evolution & Current \phantom{x} & Future \phantom{xx} \\
\midrule
Standard & $-0.008 \pm 0.07$  & $-0.001\pm 0.011$\\
Constant & $-0.1 \pm 0.3$  & $-0.001\pm 0.016$\\
Exponential & $0.01 \pm 0.05$  & $10^{-4}\pm 10^{-3}$ \\
\bottomrule
\end{tabular} 
\end{table} 

In Fig.~\ref{fig:N_Sigma} we plot $\NN$ for the three different cases. We see that the exponential evolution induces stronger deviations at early time, while in the standard and constant case deviations increase with time. Hence, measuring $\NN$ at different redshifts is necessary to capture different evolutions. We then compute the mean and uncertainty of $\Sigma_0$, from our constraints on $\NN$. The results are given in Table~\ref{tab:Sigma0}, for current and future data. Our result for the standard evolution can be compared with previous measurements of $\Sigma_0$ from DESI, DES Year 3 and Planck, which reach a precision of 0.047 (see Table 4 of~\cite{DESI:2024hhd}). Our constraints from $\NN$ are 50\% larger. This degradation is expected, since the null test depends on the derivative of $\hJ$, which the four DES measurements constrain only moderately. This is the price to pay for having a model-independent test that acts as a discriminator for models beyond $\Lambda$CDM. The goal of $\NN$ is indeed not to constrain specific parameters, such as $\Sigma_0$, but instead to capture features of gravity beyond GR. From Table~\ref{tab:Sigma0}, we see that future data will allow for much more precise measurements of the derivative of $\hJ$, which translate into significantly tighter constraints on $\Sigma_0$.    

\emph{Discussion and conclusion}: $\NN$ is a novel consistency test of GR, designed to detect any mismatch between the growth of density perturbations and that of geometric perturbations. In this work, we measured $\NN$ at four redshifts
and found no deviations in the null test. 
Stage-IV surveys are expected to increase the precision by a factor $20-50$.

While modified theories of gravity that change the propagation of light generate a non-zero null test, theories that affect only the way matter moves do not impact $\NN$. As a corollary, dark matter models involving additional interactions do not violate $\NN$. A fifth force acting on dark matter modifies indeed Euler's equation, thereby affecting both $\hJ$ and $\hf$ (via a modified $D_1$), but it preserves the relationship between them, resulting in $\NN=0$~\footnote{Note however that dark matter models violating the continuity equation, i.e.\ changing the link between $f$ and $D_1$, would break $\NN$. In models where dark matter interacts via a scalar field, such violations are typically negligible~\cite{Bonvin:2023ghp}, but this possibility should be considered if a deviation is observed.}. As such, this test is fundamentally different from the one presented in~\cite{Grimm:2025fwc}, which instead targets deviations in the dark matter sector.

The fact that $\NN$ uniquely probes the impact of gravity on light sets it apart from other gravity tests used in large-scale structure surveys. A well-known example is the $E_G$ statistic, which also combines the growth rate of structure, $f$, with gravitational lensing~\cite{Zhang:2007nk,Alam:2016qcl,Amon2018MNRAS.479.3422A, Reyes:2010tr, Blake:2016fcm, Pullen2016, Wenzl:2024sug, Abidi_2023_10419,Grimm:2024fui}. While $E_G$ deviates from its $\Lambda$CDM prediction when light propagation is modified, it is also sensitive to changes in matter propagation, since $f(z)$ -- which appears in the denominator of $E_G$ -- can deviate from its standard evolution even when light is unaffected. Furthermore, $E_G$ is sensitive to modifications of Euler's equation, such as those induced by a dark matter fifth force. In contrast, $\NN$ is sensitive only to deviations in the propagation of light, making it highly complementary to $E_G$ and a valuable addition to the suite of gravity tests.

\begin{acknowledgments}
C.B.\ and N.G.\ acknowledge support from the European Research Council (ERC) under the European Union's Horizon 2020 research and innovation program (grant agreement No.~863929; project title ``Testing the law of gravity with novel large-scale structure observables'').
\end{acknowledgments}

\clearpage                  
\onecolumngrid
            
\section*{Supplemental Materials}

\subsection*{Reconstruction of $\hJ$ and its derivative}
\label{app:reconstruction}

\begin{figure}[h!]
    \centering
    \includegraphics[width=.495\textwidth]{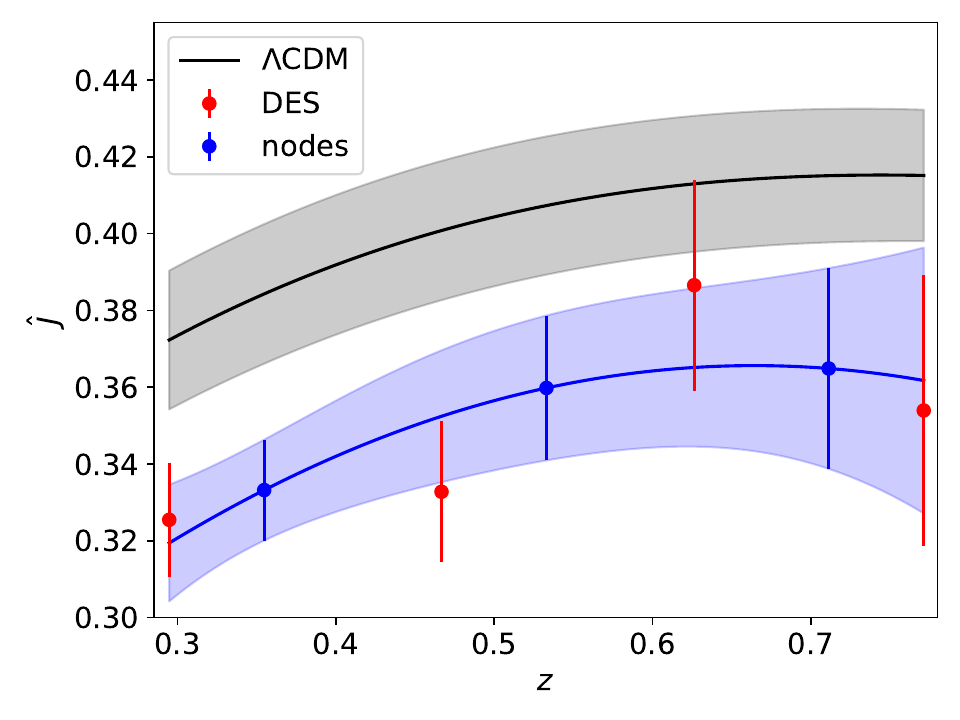}
     \includegraphics[width=.495\textwidth]{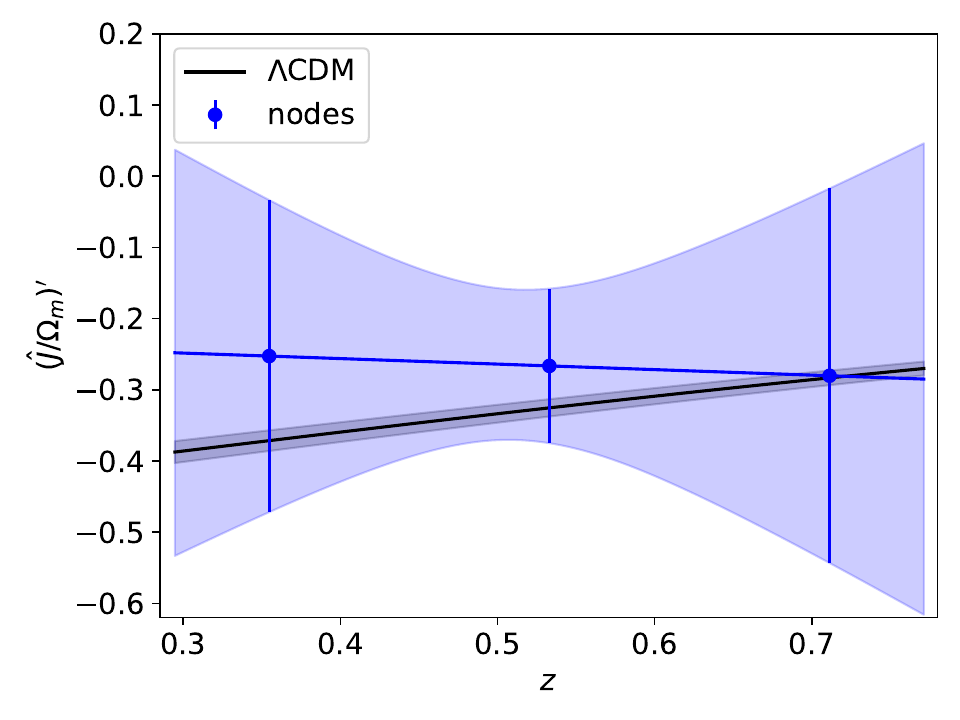}
 \caption{Reconstruction of $\hJ$ (left panel) and of $\mathrm d(\hJ/\Omega_{\rm m})/\mathrm dz$ (right panel) from three nodes. The red points are the measurements from DES with their 1$\sigma$ uncertainties. The blue points show the results at the three nodes with 1$\sigma$ uncertainties, while the blue region is the reconstruction obtained from the three nodes. The black line and gray region show the $\Lambda$CDM prediction and its 1$\sigma$ uncertainty obtained from the cosmological parameters measured in the analysis.} \label{fig:Jhat}
\end{figure}

In Fig.~\ref{fig:Jhat} we show $\hJ$ and $\frac{\mathrm d}{\mathrm dz}\left(\frac{\hJ}{\Omega_{\rm m}} \right)$ reconstructed from three nodes. We see that three nodes provide a smooth reconstruction of $\hJ$ that agrees well with the measurements from DES. Note that in this case, using cubic spline interpolation with the not-a-knot boundary conditions, the interpolated curve is a single second-order polynomial. The reconstruction is systematically below the $\Lambda$CDM prediction. This is mainly driven by the measurements in the two lowest redshift bins, which are in $2-3\sigma$ tension with $\Lambda$CDM as discussed in~\cite{Tutusaus:2023aux}. The derivative of $\hJ/\Omega_{\rm m}$ is however in good agreement with the $\Lambda$CDM prediction, as we see in the right panel of Fig.~\ref{fig:Jhat}. This is due to the fact that the reconstructed $\hJ$ has a similar slope as the $\Lambda$CDM one, and also that the uncertainty on the reconstructed derivative is quite large. Having measurements in a larger number of redshift bins will help in determining the slope more accurately. Note that the uncertainty on the $\Lambda$CDM derivative is relatively small, since in that case $\hJ$ is a known function, whose uncertainty is uniquely due to the uncertainty in the cosmological parameters. From Fig.~\ref{fig:Jhat}, we understand that having measurements of $\hJ$ in a larger number of redshift bins, from stage-IV surveys, will significantly reduce the uncertainty on the derivative of $\hJ$ and thus on the null test.

\begin{figure}[h!]
    \centering
    \includegraphics[width=.495\textwidth]{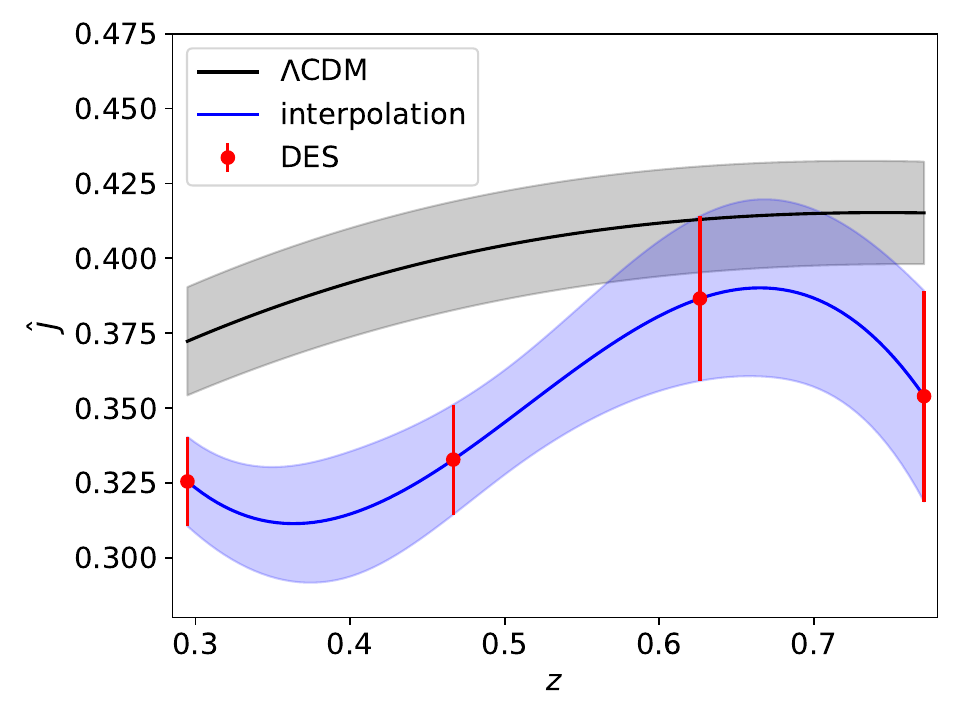}
     \includegraphics[width=.495\textwidth]{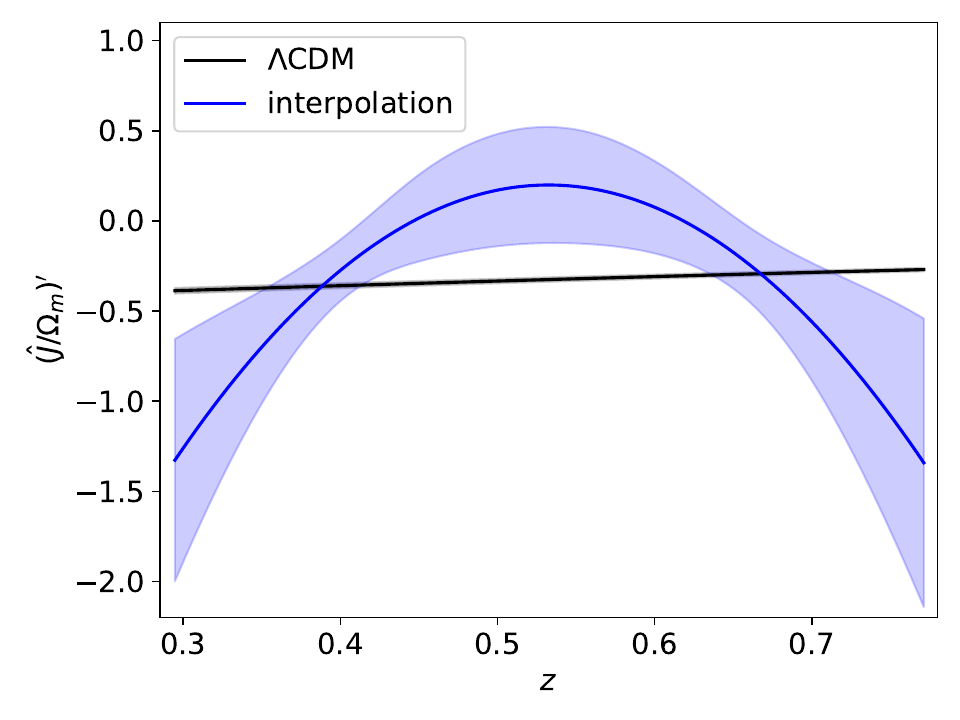}
     \caption{Reconstruction of $\hJ$ (left panel) and of $\mathrm d(\hJ/\Omega_{\rm m})/\mathrm dz$ (right panel) from interpolation. The red points are the measurements from DES with their 1$\sigma$ uncertainties. The blue region is the reconstruction obtained from interpolation. The black line and gray region show the $\Lambda$CDM prediction and its 1$\sigma$ uncertainty obtained from the cosmological parameters measured in the analysis.} \label{fig:interpolation}
\end{figure}

\begin{figure}[h!]
    \centering
    \includegraphics[width=.495\textwidth]{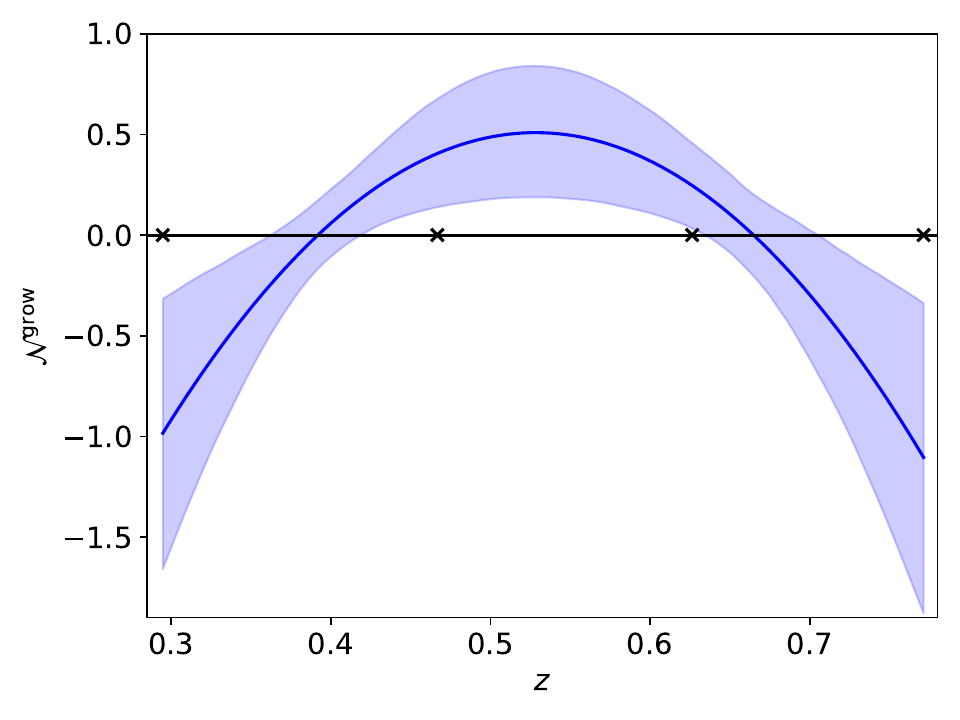}
    \caption{Measurements of $\NN$ from current data sets,  together with the 1$\sigma$ uncertainties, obtained from interpolation of $\hJ$. We show the reconstructed $\NN$ over the whole redshift range. The black crosses indicate the position of the DES redshifts where $\hJ$ is measured.} \label{fig:Ninterpolation}
\end{figure}

We then explore an alternative method, where instead of choosing a number of redshift nodes and optimizing the values of $\hJ$ and $\frac{\mathrm d}{\mathrm dz}\left(\frac{\hJ}{\Omega_{\rm m}} \right)$ at each step of the MCMC chain, we interpolate between the four measurements at each step (using cubic spline interpolation). In Fig.~\ref{fig:interpolation} we show the resulting $\hJ$ and the derivative. We see that in this case $\hJ$ oscillates, and is thus more consistent with the $\Lambda$CDM prediction at high redshift. This results in a derivative that changes sign, as we see in the right panel of Fig.~\ref{fig:interpolation}. $\NN$ is plotted in Fig.~\ref{fig:Ninterpolation} and it inherits the change of sign from the derivative.

We see that also in this case, $\NN$ is broadly consistent with zero (at $1.5\sigma$). The uncertainties are slightly larger than in the three nodes case. In particular, at $z_{\rm nodes}$ we obtain $\sigma_{\NN}\in \{0.28,0.32,0.37 \}$. We find that this interpolation method gives exactly the same result as the nodes method, in the case where we choose four nodes. Indeed, with four nodes there is enough freedom for the interpolated curve to pass through the four data points at each step of the chain, leading to the same reconstruction of $\hJ$, its derivative and $\NN$. From this comparison of the two methods, we find that our conclusion that no deviations from GR can currently be detected through $\NN$ is robust.  
The uncertainty over the redshift range is also quite robust, but the particular value of the uncertainty at specific redshifts does depend on the method. The choice of three nodes provides tighter constraints than the interpolating method, which is effectively the same as choosing four nodes. This is expected, since with three nodes we reconstruct $\hJ$ and its derivative with a smaller number of free parameters, which are therefore better constrained.

\subsection*{Additional tables}
\label{app:tables}

In Table~\ref{tab:Ncurrent} we list the values of $\NN$ with the 1$\sigma$ uncertainties obtained from current data sets. These values are plotted in Fig.~\ref{fig:current}. In Table~\ref{tab:Nfuture} we list the values of $\NN$ with the 1$\sigma$ uncertainties forecasted for DESI and LSST. These values are plotted in Fig.~\ref{fig:future}.

\setlength{\tabcolsep}{0.8em} 
\begin{table}[h!]
\centering
\caption{Mean and 1$\sigma$ uncertainties for $\NN$ obtained from current data sets.} \label{tab:Ncurrent}
\begin{tabular}{c r} 
\toprule
 $z$ & $\NN$\phantom{x}\\
\midrule
0.35  & $0.08 \pm 0.22$  \\
0.53   & $ 0.04 \pm 0.11 $ \\
0.71   &$ -0.01 \pm 0.26 $  \\
\bottomrule
\end{tabular} 
\end{table}

\begin{table}[h!]
\centering
\caption{Forecasted mean and 1$\sigma$ uncertainties for $\NN$, combining DESI with LSST.} \label{tab:Nfuture}
\begin{tabular}{@{}c r@{}}
\toprule
$z$ & $\NN$\phantom{xxx} \\
\midrule$0.05$   &  $-0.002\pm 0.059$ \\
$0.15$   &  $0.002\pm 0.024$ \\
$0.25$   &  $0.004\pm 0.015$ \\
$0.35$   &  $0.003\pm 0.012$ \\
$0.45$   &  $0.002\pm 0.014$ \\
$0.65$   &  $-0.0005\pm 0.0070$ \\
$0.75$   &  $-0.0012\pm 0.0055$ \\
$0.85$   &  $-0.0012\pm 0.0049$ \\
$0.95$   &  $-0.0004\pm 0.0044$ \\
$1.05$   &  $0.0010\pm 0.0045$ \\
$1.15$   &  $0.0019\pm 0.0045$ \\
$1.25$   &  $0.0019\pm 0.0040$ \\
$1.35$   &  $0.0010\pm 0.0044$ \\
$1.45$   &  $0.0001\pm 0.0053$ \\
$1.55$   &  $-0.0004\pm 0.0059$ \\
$1.65$   &  $-0.0007\pm 0.0068$ \\
$1.75$   &  $-0.0006\pm 0.0099$ \\
$1.85$   &  $-0.001\pm 0.011$ \\
\bottomrule
\end{tabular}
\end{table}

\subsection*{$\mu-\Sigma$ parametrization}
\label{app:mueta}

One well-studied modification of GR consists in phenomenologically modifying Poisson's equation through a function $\mu$ and introducing a gravitational slip $\eta$:  
\begin{align}
k^2\Psi(k,z)&=-4\pi G\mu(z)\delta\rho(k,z)\, ,\label{eq:mu}\\  
\Phi(k,z)&=\eta(z)\Psi(k,z)\, . \label{eq:eta}
\end{align}
In full generality $\mu$ and $\eta$ can also depend on $k$. Since in many theories of gravity in the quasi-static approximation, the dependence is very weak and can be neglected~\cite{Gleyzes:2015pma,Gleyzes:2015rua,Raveri:2021dbu}, we apply this simplification here. Note that allowing for a dependence in $k$ would require to redo the measurements of $\hf$ and $\hJ$ also allowing for such a dependence.

Einstein's equations are supplemented by Euler's equation, which governs the motion of galaxies
\begin{align}
V'(k,z)+V(k,z)=\frac{k}{\HH}\Psi(k,z)\, , \label{eq:Euler}
\end{align}
where a prime denotes a derivative with respect to $\ln a$. Combining Eqs.~\eqref{eq:mu}-\eqref{eq:Euler} we can write
\begin{align}
&\Psi_W(k,z)=-\frac{3}{2}\left(\frac{\HH(z)}{k}\right)^2\Omega_{\rm m}(z)\Sigma(z)\delta(k,z)\, , \label{eq:Psi_Sigma_bis}\\  
&V'(k,z)+V(k,z)=-\frac{3}{2}\frac{\HH(z)}{k}\Omega_{\rm m}(z)\mu(z)\delta(k,z)\, . \label{eq:Euler_mu}
\end{align} 
We see therefore that $\Sigma\neq 1$ generates a change in the Weyl potential, i.e. a change in the way light propagates. On the other hand $\mu\neq 1$ changes the way matter moves. Combining Eq.~\eqref{eq:Euler_mu} with the continuity equation, we obtain a second-order evolution equation for $\delta$:
\begin{align}
\delta''+ \left(1+\frac{\HH'}{\HH}\right) \delta'-\frac{3}{2}\frac{\Omega_{m,0}}{a}\left(\frac{\HH_0}{\HH}\right)^2 \mu\,\delta=0\,. \label{eq:deltaevol_mod}
\end{align}
Hence $\mu$ directly affects the growth function $D_1$.

\bibliography{nulltest_Jhat}
\bibliographystyle{apsrev4-1}

\end{document}